\documentclass[11pt,twoside,A4]{article}
\usepackage{times,fancyhdr}
\usepackage[dvips]{graphicx}
\usepackage{latexsym}
\usepackage[affil-it]{authblk}
\usepackage{mathptm}
\usepackage{amsmath}
\usepackage{amssymb}
\usepackage{color}
\usepackage{setspace}
\usepackage{hyperref}
\usepackage[top=4cm, bottom=4cm, left=3.8cm, right=3.8cm]{geometry}

\def\beq{\begin{equation}}
\def\eeq{\end{equation}}
\def\beqn{\begin{eqnarray}}
\def\eeqn{\end{eqnarray}}

\newcommand{\be}{\begin{equation}}
\newcommand{\ee}{\end{equation}}
\newcommand{\bea}{\begin{eqnarray}}
\newcommand{\eea}{\end{eqnarray}}

\begin{document}

\title{Screams for Explanation:\\ Finetuning and Naturalness in the Foundations of Physics 
}
\author{S.\ Hossenfelder}
\affil{\small Frankfurt Institute for Advanced Studies\\
Ruth-Moufang-Str. 1,
D-60438 Frankfurt am Main, Germany
}
\date{}
\maketitle
\begin{abstract}
 
We critically analyze the rationale of arguments from finetuning and naturalness in particle physics and cosmology. 
Some other numerological coincidences are also discussed.
\end{abstract}

\section{Introduction}

The scientific method, so the idea, is a virtuous cycle of hypotheses-generation followed by experimental test and subsequent evaluation. But where do hypotheses come from to begin with? 

Scientists do not randomly guess hypotheses -- that would waste too much time. Instead, much of the scientific enterprise today is dedicated not to testing hypotheses but to selecting hypotheses worthy of test. All through their education, scientists learn to identify worthwhile research topics and then judge their own and colleagues' work by the so-acquired experience. 

In this paper, I will analyze whether ``unnaturally'' small or large numbers require explanation and thus whether hypotheses that explain such numbers are promising research topics. The answer I shall offer is ``sometimes.''  An unnatural number requires explanation when it is in a quantifiable sense unlikely. Unfortunately, as I will demonstrate here, most unnatural numbers presently studied in the foundations of physics are not quantifiably unlikely. It follows that the corresponding problems of naturalness are ill-defined and might not be problems at all.

In the next section I classify different types of finetuning. Sections \ref{cosm} and \ref{part} briefly summarize the ``big problems'' under discussion here. The reader familiar with these topics may skip to section \ref{prob} in which I lay out the shortcomings of finetuning arguments.  Section \ref{faq} is a forward-defense against frequently asked questions. Section \ref{Why} explains the relevance of the present study. I summarize in section \ref{conc}. 

\section{Types of Finetuning}
\label{types}

A good hypothesis must, most importantly, be compatible with already existing knowledge, including both data and requirements of mathematical consistency.\footnote{Here and in the following ``consistency'' refers to the absence of internal contradictions in a theory's regime of applicability. It does {\sl not} imply that a theory has empirical support.}  To be interesting, the hypothesis must moreover make plausible new predictions though such predictions may be not be amenable to test in the near future.\footnote{This raises the question how much effort should be made to work out details of theories that have not yet been tested, but this is a question which shall not concern us here.} 

In the foundations of physics -- on which this paper focusses -- the requirement that a hypothesis be mathematically consistent, compatible with existing data and still make a new prediction is difficult to fulfil. Since physics is a mature discipline, current theories work extremely well already and are therefore hard to improve further. There isn't even much need for improvement, because the theories that currently constitute the foundations of physics -- the cosmological concordance model and the standard model of particle physics - -explain the presently available data just fine.

So what is a theorist in the foundations of physics to do? A cursory scan of the literature reveals that many of the research-efforts in theoretical high energy physics and cosmology focus on a few ``big problems.'' In this paper, we will have a closer look at these problems and investigate how problematic they really are.

We will not discuss here the history of arguments from naturalness and finetuning because this has been done before. The reader interested in the historical development is referred to \cite{1,2}. A previous survey of finetuning arguments can be found in \cite{3}  and the unnaturally small mass of the Higgs-boson in particular was discussed in \cite{4}.

\subsection{Naturalness (in general) }

Physicists use the word ``natural'' in general to mean that a theory's dimensionless parameters (ie those without units) are not much larger or much smaller than 1. Since any small number can be converted into a large number by taking its inverse, these two cases do not have to be distinguished from each other.  

Physicists usually do not quantify exactly how much larger or smaller than 1 a number may be. The tolerance for how far away from 1 is permissible depends on any individual's belief that such a number may derive from an as-yet undiscovered calculation. This belief is strongly based on experience. By experience, for example, it is not difficult to obtain factors of about 100 from powers of $2\pi$ that frequently appear in volume integrals. 

It follows from the definition of naturalness that two numbers which are much closer together than each number's absolute value are also suspicious and ``unnatural.'' That is because in this case the difference between the two numbers would be a small number. I mention this example because naturalness problems in physics often originate in such small differences.

The belief in naturalness is usually rationalized by claiming that numbers which are very large or very small are unlikely. They appear cherry-picked for no other
reason than correctly describing an observation and require further explanation.

\subsection{Technical naturalness (in particular)}

The notion of ``technical naturalness'' applies only to quantum field theories in particular. Technical naturalness, originally formulated by `t Hooft \cite{5}, is a weaker criterion than naturalness in general because it still permits certain small numbers. A small number is permitted if it has an explanation, typically because it is protected by a symmetry. 

To understand technical naturalness, first recall that quantum field theories are energy-dependent. At higher energies, new processes become possible, and interactions can decrease or increase in relevance. This means that the dimensionless numbers (``parameters'') which appear in a quantum field theory depend on energy; the ``parameters run'' as the terminology has it. The energy in question is determined by the type of experiment for which one wants to make predictions and the running of parameters can be calculated using the renormalization group equations \cite{6}.  

The best known case of running parameters are the standard model's coupling constants which set the strengths of the three interactions. The coupling constant of the strong nuclear force, for example, becomes smaller at higher energies, a property known as ``asymptotic freedom.'' The other couplings also run with energy. 

It follows that if we change the energy at which a theory is applied, the theory will trace out a curve in an abstract ``theory-space.'' In this theory-space each point represents a theory or, since the set of parameters for all possible interaction terms in the Lagrangian defines the theory, each point is a combination of parameters respectively. The change of all possible theories with energy is known as the ``flow'' in theory-space.\footnote{In practice one does not, of course, study the space of all possible quantum field theories, but only that of certain classes of theories, typically chosen by field-content and symmetry-requirements.} 

To understand technical naturalness, note now that each theory at high energies is connected by the renormalization group flow to a theory at low energies and vice versa. This makes it meaningful to ask what happens with the theory at low energies if we change the corresponding parameters of the theory at high energies because the two parameter-sets are related by the renormalization group equations. 

To digest the literature on the subject, it is helpful to know that particle physicists often refer to high energies as ``ultraviolet'' (UV) and to low energies as ``infrared'' (IR). A theory is then said to be technically natural if the theory in the IR does not sensitively depend on the choice of parameters in the UV.

Phrasing naturalness in terms of a sensitive dependence on the theory at high energies is useful because it allows physicists to quantify just how unnatural a theory is. Several different measures for this have been introduced in the literature, but the exact definitions are not so relevant here. For our purposes it is more relevant to understand the underlying reasoning for why these measures quantify something of interest. 

High energies correspond to high resolution and hence short distances. It is therefore basic reductionist reasoning that the theory at high energies is more fundamental. The idea of quantifying naturalness through the sensitivity on the high-energy parameters is then that the compatibility with what we observe at low energies should not require improbable coincidences in the more fundamental theory. Because we don't know any better, so the argument, the parameters at high energies could really have had any values, and the precise choice should not affect what happens at low energies where we find the standard model. In this way, technical naturalness reflects the idea of general naturalness, that humans shouldn't cherry-pick parameters. 

A theory, then, is technically natural if getting a low-energy phenomenology that is within measurement precision of the standard model is likely if we were to randomly pick the parameters at high energies. This assumption is ill-defined because we don't have a probability distribution on parameter space, neither at low energies nor at high energies. I explain which problems this brings in section \ref{prob}. But first some more words on technical naturalness.

I often hear technical naturalness being referred to as ``the UV physics decouples.'' I have found this phrase to cause much confusion, so allow me some words of caution. 

The change of parameters in the UV, which is done to quantify technical naturalness, is not a process that is physically possible. The theory that describes our world is defined by one particular choice of parameters. (The reader who has trouble understanding the meaning of these parameters may keep in mind the example of the standard model's coupling constants.) At a given energy, these parameters have some specific values. We can't change these values because that would amount to changing the laws of nature.

The physics at high energies indeed ``decouples'' (for all we know) but that alone is a not a sufficient criterion for a natural theory. That the physically possible processes at high energies decouple at low energies means for example that to calculate the orbit of the moon you don't need to know what the electrons in the moon's atoms do. Nobody knows why that is so, but this decoupling is evidently a property of nature. Decoupling is necessary to use effective field theory and is hence an assumption that underlies the whole framework of the renormalization group already \cite{7}.  This means the UV physics decouples whenever effective field theory can be used, regardless of whether or not the theory is natural. 

Technical naturalness, therefore, is a criterion separate from the decoupling of short-distance processes. It quantifies a sensitivity to a virtual (``mathematical'') change of parameters, not a sensitivity to a change that can actually happen. 

Another common confusion is that naturalness is necessary for the validity of making an expansion of the theory because the convergence of higher order contributions assumes that the parameters in the expansion don't unduly increase. True, but again, while this is certainly an assumption necessary to make sense of the whole framework on which naturalness builds, it is not a sufficient (or even necessary) requirement for naturalness.  This can be illustrated by the typical naturalness problems which are not so much worries about the overall magnitude of parameters (relevant for the expansion) but about ``suspicious'' cancellations between them. (Indeed such cancellations might help a series converge.) 

So why believe in naturalness? I am not sure just why naturalness has become such an exceedingly popular criterion to decide whether or not a theory is promising. A key reason is certainly that the masses of particles in the standard model are all technically natural, except for the Higgs-boson \cite{8}.  

There are further three commonly named historical examples in which the presence of an unnatural number signaled that a theory had reached its limits and some new effect had to show up beyond a certain energy. These three examples (see \cite{2}  for details) are: 1) the mass of the electron is small compared to the contribution that stems from the self-energy of its electric field (technically infinitely large), 2) the difference between the masses of the two charged pions is much smaller than either one's mass, and 3) the absence of flavor changing neutral currents in the standard model, which signals that a constant in front of a term enabling such processes must be small. 

The first two examples were found to be naturalness problems only after the new processes had been observed (the positron and $\rho$-meson, respectively). The third
example lead to the prediction of the charm quark, to my knowledge the only prediction ever made based on a naturalness-argument \cite{charm}.

On the other hand, we now know of at least three failures of naturalness: The cosmological constant, the mass of the Higgs boson, and the strong CP problem. These will be discussed below. 

(Note that the decoupling theorem \cite{9}, despite its name, does not prove technical naturalness. Besides using an outdated regularization scheme it assumes perturbative renormalizability of the underlying theory which isn't only a questionable assumption in general, we know it to be not correct in the case of gravity.)

\subsection{Anthropic Finetuning }

Finetuning is also often used to quantify how much the parameters in our theories could be changed so that life as we know it would still be possible. I will not discuss this case here because it rests on an entirely different logic.

\section{Finetuning In Cosmology}
\label{cosm}

\subsection{The Cosmological Constant Problem}

The universe expands. And not only that, according to presently available data its expansion is speeding up by the day. Trouble is, in the framework of general relativity an accelerated expansion cannot be caused by matter or energy of any type that we know. 

An accelerated expansion of the universe requires a peculiar type of ``dark'' energy to get the expansion to accelerate. This energy must grow in proportion to the volume of the universe, or at least do something very similar to this. To phrase it differently, the density of dark energy must remain constant as the universe expands.

The simplest type of dark energy is just a constant, known as the cosmological constant and usually denoted $\Lambda$. Long believed to zero, the best current data put $\Lambda$ at a small but nonzero, positive, value. 

The cosmological constant has units, so what it means for it to be small requires explanation. In quantum field theory, the vacuum carries a non-zero energy-density which derives from virtual particle contribution. For the standard model, the dominant contribution to the vacuum-energy -- let us call it $\lambda$ -- is proportional to the fourth power of the mass of the heaviest particle \cite{10}. The heaviest currently known particle is the top quark, which has a mass of about $10^{11}$eV.\footnote{In units in which the speed of light and Planck's constant are equal to 1.}  The energy-scale associated with the cosmological constant is about 1/10 eV. The ratio between the two energies is about $10^{-12}$. The unknown origin of this small number is the cosmological constant problem.\footnote{Or at least the problem we will discuss here. The literature distinguishes several cosmological constant problems, but the other ones aren't so relevant for what follows.}
  
However, the contribution from quantum field theory is not in and by itself observable. In observables, this contribution always appears together with a free constant from general relativity. Let us call this other constant (sometimes referred to as the ``bare'' cosmological constant) $\lambda'$.  It can be chosen by the requirement that $\lambda+\lambda' = \Lambda$, ie that the two contributions together reproduce the measured value, $\Lambda$. This requirement follows the same logic by which infinities in quantum field theories can be removed: By noting that the infinities are not themselves observable and therefore it is possible to subtract another, suitably chosen, infinity so that a finite term remains whose value is then determined by measurement. The same is possible for the cosmological constant, except that here the subtracted term is (usually assumed to be) finite. 

However, since $\Lambda$ (the observed value) is much smaller than $\lambda$ (the contribution from the standard model), the requirement to reproduce the observed value means that $\lambda'$ (the bare contribution from general relativity) and $\lambda'$ must almost, but not exactly cancel. These constants are dimensionful, hence speaking about their absolute values is meaningless. But the statement can be rewritten without units to say that $1+\lambda/\lambda'$ must be a small number. 

It is here where finetuning arguments become relevant. The typical argument goes like this: We don't know $\lambda$, so we will assume that it could take on any value between $- m_{\rm p}^4$ and $m_{\rm p}^4$ where $m_{\rm p}$ is the Planck mass and approximately $10^{29}$eV. If the value of the constant is randomly chosen with uniform probability in that interval, then the probability that it will just by chance almost cancel $\lambda'$ and leave behind $\lambda$ is miniscule. 

Since the cosmological constant scales with the fourth power of mass, it is highly sensitive to whatever is the corresponding parameter at high energies. For this reason the above quoted value for the vacuum energy from the standard model -- which scales with the mass of the heaviest particle--has a theoretical uncertainty that is itself estimated to scale with the Planck mass. If there are any heavier particles that we have not yet seen, for example, these would dominate the contribution. That's why the problem is sometimes stated in terms of comparing the Planck density to the (density associated with) the cosmological constant, which results in the (more frequently quoted) 120 orders of magnitude mismatch. 

However, regardless of exactly how one formulates the problem, the cosmological constant is both generally and technically unnatural.

\subsection{The Flatness Problem}

The spacial curvature of the universe is presently very small, so small that it's compatible with a flat universe. The contribution of curvature to the expansion of the universe (the Friedmann equations) however increases relative to the contribution from matter. To see this, divide the first Friedmann equation through the square of the Hubble-rate to get a dimensionless expression:

\beqn
\frac{8\pi}{3 m_{\rm p}^2} \left( \frac{\rho_0^{\rm r}}{(a \dot a)^2} + \frac{\rho_0^{\rm m}}{a \dot a^2} \right) + \frac{\Lambda}{3}  \frac{a^2}{\dot a^2} - \frac{k}{\dot a^2} =1
\eeqn

Here, $a$ is the scale-factor, $\rho_0^{\rm r}$ and $\rho_0^{\rm m}$ are some initial values for the density of radiation and matter, respectively, and $k$ is the (dimensionful) curvature parameter. Since $a$ increases, the cosmological constant term will eventually come to dominate. But also the contribution from the curvature term grows faster than the contribution from both radiation and matter. This means if the contribution from the curvature density is unobservably small today, in the past it must have been tiny compared to the other densities. Where does its small initial value come from? That's the flatness problem. 

Again the expectation is that ``typical'' numbers should be of order one, while for the universe to be flat, one needs a factor of $10^{-60}$ or so to get today's value to be compatible with observation (the exact value depends on the time when initial conditions were set and is not so relevant in the following). 

The flatness problem is one of the problems that the theory of inflation -- the idea that the universe underwent a phase of exponential expansion -- attempts to solve \cite{11}.  The supposed problem can be removed by choosing a value that's compatible with observation. Statements about the value's likelihood cannot be made because the probability distribution is unknown and cannot be empirically determined because we have only one universe in the sample. 

\subsection{The WIMP Miracle}

Weakly interacting massive particles ({\sc{WIMP}}s) are one of the most popular candidates for dark matter. Their popularity derives from a numerical coincidence, which is that particles with mass nearby the electroweak scale and with a cross-section typical for the weak interaction would be formed in the early universe with about the right abundance for dark matter. Unfortunately the particle have not been detected. The expected cross-section has been repeatedly revised to stay below experimental bounds, see eg  \cite{12}.

\subsection{Other Finetuning Problems in Cosmology}

There are a variety of other finetuning problems in cosmology which we will not discuss here because while space and time might be infinite in general, mine in particular aren't. The omitted problems include: The horizon problem, the monopole problem, the coincidence problem of the cosmological constant, the baryon asymmetry, and several finetuning problems that appear in inflationary models. 

\section{Finetuning in Particle Physics}
\label{part}
\subsection{The Higgs Mass}

The Higgs is the only fundamental scalar field. For this reason, the mass of the Higgs-boson receives large contributions from loop corrections\footnote{These corrections are often said to be due to ``quantum fluctuation,'' which is not wrong but sometimes causes confusion by bringing up the question just what is fluctuating and why. Suffices to say that this is just a word given to some terms in an equation.}, a problem which does not occur for any other particle. These contributions are estimated to be of the order of the energy where the theory (in this case, the standard model) breaks down, which is close by either the Planck mass or the energy scale where a grand unified symmetry is (believed to be) restored. This means the contribution from the loop corrections is at least 13 orders of magnitude larger than the actual mass of the Higgs-boson.

This problem can be remedied by subtracting the (total) contribution from all quantum fluctuations and henceforth ignore it because it is not in and by itself an observable. That the quantum contributions for the Higgs-boson's mass don't make practical trouble is evidenced by the large number of excellent predictions which agree with measurements in spite of the supposed problem.\footnote{In many cases the question whether a problem really is a problem can be answered by observing what economists refer to as ``revealed preferences.'' It means, in brief, look at what they do, don't listen to what they say. I think this criterion is of great use also in theoretical physics. }  

However, the introduction of a new term that almost but not exactly cancels the contribution from the quantum fluctuations is thought to be fine-tuned for the same reason that the cosmological constant is fine-tuned. If the two contributions had a typical, almost uniform distribution over an interval from minus the Planck mass to the Planck mass, then the probability that they almost cancel is tiny. 

The most popular solution to remedy the unnaturalness of the Higgs-boson's mass was supersymmetry. In supersymmetric extensions of the standard model, additional particles appear beyond some energy scale. Supersymmetry renders the Higgs-boson's mass natural because it enforces a cancellation between different contributions to the mass. 

However, the leading contributions to the Higgs-boson's mass then scale with the masses of supersymmetric partners. This means if the supersymmetric particles are heavier than the Higgs-boson itself, then the naturalness problems return. The same happens for any other type of new physics that comes in at some energy scale which must be beyond what we probed so far. This is why the data delivered by the LHC has now ruled out a technically natural explanation for the Higgs mass. Though I want to add that of course it is still possible a natural solution exists, just that it is more complicated than previously thought.

\subsection{The Strong CP problem}

Quantum-electrodynamics is symmetric under a CP-transformation, which is a combination of changing the electric charge of a particle to its opposite (C) and changing the parity of the particle (P). The weak nuclear force violates this symmetry, as we have known since the 1960s. The strong nuclear force could violate it but for unknown reasons doesn't -- at least no such symmetry violation has been seen in any experiments. 

This can be formalized by writing down a contribution to the Lagrangian that violates the symmetry and saying that the factor in front of it -- the theta-parameter, denoted 
$\theta$-- is either very small or zero. Why this factor is so small is known as the ``strong CP problem.''

A solution to the strong CP problem is to turn the theta-parameter into a dynamical field that takes on a minimal value in a self-induced potential. This solution works and is technically natural, but it was noticed quickly after its proposal that the field would be accompanied by a particle, the ``axion'' \cite{axion,axion2}. The axion was looked for and experimentally ruled out \cite{noaxion,noaxion2}.  

In response to this, the axion models were amended so that the axion became harder to detect. Dubbed the ``invisible axion,'' it has become one of the most popular candidates for dark matter, though there is still no evidence for its existence.\footnote{The reader be warned that what was once called ``invisible axion'' is now in the literature often referred to as just ``axion.'' }

\subsection{Gauge Coupling Unification}

As we saw earlier, the three coupling constants of the standard model are energy-dependent. Their energy-dependence is so that they converge towards each other. In the standard model, however, the curves do not ultimately meet in one point. If one adds particles to the standard model which are so heavy that we have not yet seen them, this changes the slopes of the running of the coupling constants. Models in which the three curves meet in one point are said to allow for ``gauge coupling unification.''

Gauge coupling unification is interesting because if the three interactions of the standard model arise from one unified theory with a symmetry that was broken at high energies, then the gauge couplings should meet at an energy close to the breaking scale. But while this is an appealing idea, it is neither necessary for consistency nor supported by data. 

It has been known since the 1990s that if one adds supersymmetric partner particles to the standard model, then the running gauge couplings happen to meet in one point (up to measurement precision) \cite{13}.  This numerical coincidence has been one of the biggest motivators for supersymmetry. The argument is that such a meeting of curves would be unlikely to happen by chance and is hence requires an explanation, like for example supersymmetry. 

However, how well the gauge couplings meet in supersymmetric models depends on the masses of the superpartners. Since the new data from the LHC has pushed up the lower bounds, the superpartners -- if they exist -- must now be quite heavy, and this makes gauge coupling unification worse. For this reason, some theorists now argue that additional terms are relevant in the calculation so that gauge coupling unification can still be maintained in supersymmetric models (presumably implying that the couplings shouldn't have met without these terms to begin with) \cite{14}.  

\section{Problems with Finetuning Arguments}
\label{prob}

\subsection{Circularity}

The major problem with finetuning arguments both in cosmology and in particle physics is the reference to probabilities without defining a probability distribution, or, if a distribution is defined, the failure to explain where the distribution itself comes from. 

It is commonly -- most often without stating explicitly -- assumed that the probability distribution is almost uniform over an interval that (for the dimensionless parameters) stretches from -1 to +1. A prototypical example is a normalized Gaussian of width 1 around 0, which the reader may keep in mind as example for the following. This means one assumes a width of order 1 to justify that a probable parameter is of order 1, which is an obviously circular argument.

It is easy to see that the argument is circular because one could, eg, assume a probability distribution with a width of, say, $10^{-14}$, which would lead to the conclusion that the ``typical'' difference between two randomly picked numbers is of order  $10^{-14}$. We get out what we put in. On that account, hence, any number is equally ``natural.'' 

A particle physicist would likely object at this point that a probability distribution which explicitly refers to a small number like  $10^{-14}$ is itself already finetuned. But this merely brings up the question what's the probability of a probability distribution and so on, resulting in an infinite regress unless some number or distribution is just postulated to be better than all others. 

If one wants to remove the problem of circularity one necessarily has to postulate a probability distribution which brings back exactly the arbitrary choice that the criterion of naturalness was supposed to remove. Naturalness is hence either ill-defined or meaningless.

This need to choose a probability distribution for measures of technical naturalness used to be well-known. In 1995, Anderson and Castano, in one of the first papers to quantify technical naturalness, clearly stated that the choice of a probability distribution ``necessarily introduces an element of arbitrariness to the construction'' \cite{15} But the issue is no longer discussed in today's literature.

\subsection{Occam's Razor}

A probability distribution from which to calculate the most likely choice of parameter adds unnecessary structure to the theory and is thus in conflict with the dictum of simplicity. We could have chosen a parameter and be done with it. The probability distribution and all the not-observed values of the parameters are unnecessary for the derivation of any observable and they should therefore be stripped by Occam's razor. 

Indeed, as I noted earlier, this is evident by just looking at how practitioners in the field do their calculations. No one in their right mind would start with defining a useless probability distribution over a space from which eventually only one value is needed. 

The anthropic prediction for the cosmological constant is often named as an example for the usefulness of probabilistic arguments \cite{16}.  This argument amounts to guessing a probability distribution, then adding anthropic priors, and then deriving a likelihood for our observation. It is certainly interesting that the so-obtained most likely value agrees well with actual measurements. From an axiomatic standpoint, however, this calculation merely replaces guessing what we observe by guessing a probability distribution for what we observe. It's still a guess, albeit one that can't be ruled out because it's probabilistic anyway.

In recent years, arguments using Bayesian inference have become fashionable, both in cosmology and in particle physics. Some people seem to believe that this changes anything about the problems with naturalness, but Bayesian inference just moves the problem from the choice of a probability distribution to the choice of priors. The priors are then assumed to be ``natural'' to justify what is natural (see also section \ref{faq}). 

But it doesn't matter which way one attempts to calculate probabilities, this will not put naturalness arguments on a solid mathematical footing. The reason is that, when it comes to the laws of nature, we don't observe repeated events or sample over many outcomes. Any talk about probability distributions or priors refers to the distribution of theories in some mathematical space, almost all of which we cannot observe. We have only one set of laws of nature.

The multiverse is the assumption that all of these unobservable theories are as real as ours. In the case of the multiverse at least the problem of calculating probabilities is widely acknowledged and has entered the literature under the name ``measure problem.'' The attempt here is to calculate a measure\footnote{A ``measure'' is what's necessary to assign weights to the elements of a set. For all practical purposes it's the same as a probability distribution.} according to which the parameters that we observe in our universe are (ideally) the most likely ones (compatible with the existence of life). 

One can only hope that such a measure would not require more parameters than the parameters that one can supposedly calculate with it. But even so, imagine that approach was one day successful and someone would indeed manage to find a measure according to which the values of parameters we observe are the most likely ones, using fewer parameters as input. It would mean that physicists had discovered a way to calculate (some of) the parameters of current theories using a simpler set of parameters by searching for some optimum of some function. This would be great, but all the talk about probability distributions could be removed from this finding.

I want to emphasize that I do not say one should not pursue such thoughts. It seems possible to me that reformulating problems in terms of probability distributions on a multiverse will help with finding a solution. Insights sometimes come in unexpected ways. I just want to make clear that doing so cannot be justified on rational grounds.

\subsection{No ``Finetuning,'' No Theory}

Finetuning problems arise from an attempt to quantify the probability of some specific assumptions of physical theories, that being the numeric values of dimensionless parameters. But all our theories have many other assumptions that are chosen for the only purpose of explaining observations. 

General relativity, for example, postulates that we live in a Riemannian manifold, and quantum mechanics postulates states are described by vectors in a Hilbert-space (or Fock space, respectively), and the axioms of Hilbert spaces, and so on. We also postulate, for no particular reason other than that it describes what we see that observables are real-valued, that vacua are stable, and that infinities aren't physical. None of these assumptions are mathematically necessary. They're there just because they work.

We could now start discussing what's the probability to get any set of assumptions which we use from out of the infinite number of mathematically consistent sets of axioms that we could have picked. But we don't discuss this. That's because the purpose of science is do describe observations and we simply pick those assumptions which are up to the task. Why make an exception for numbers?

As an aside, this misunderstanding of the purpose of a scientific explanation is the origin of most types of multiverses. They arise because some physicists refuse to select assumption ``just because'' they describe observations.

\subsection{Ambiguous Parameters (Technical Naturalness in Particular)}

Any quantification of technical naturalness uses a specific set of parameters in theory-space, which are chosen by a certain basis in the expansion of the Lagrangian. A different choice of basis in theory-space can be used to remove naturalness problems by suitably expanding or shrinking certain sectors of the parameters. Of course one could then complain that such a choice of basis wouldn't be natural, but that brings on the question which -- if any -- basis is natural and what that even means.  

\subsection{Technical Naturalness hides Finetuning}

Let us now consider that we had a probability distribution at high energies, so that we could quantify finetuning. Even so, that the standard model without the Higgs is natural doesn't mean that the theory at high energies is not finetuned. It means that the standard model isn't sensitive to whether or not the theory at high energies was finetuned. 

This was not, of course, the reason why naturalness was introduced to begin with. But in hindsight, that the masses of the standard model particles are natural besides that of the Higgs-boson means the Higgs-boson is the only particle that allows us to decide whether or not the underlying theory is finetuned. Such a conclusion could not have been drawn from the other masses to begin with, hence projecting it on the Higgs boson wasn't a rational inference.  

Let me emphasize that the argument here is not that it is wrong to think the theory at low energies is sensitive to the parameters at high energies. This sensitivity is a property of the theory and I do not question it. I am questioning the reason to think that this sensitivity matters or, more to the point, that the absence of such sensitivity means a theory is a promising explanation for natural phenomena. 

\section{Frequently Asked Questions}
\label{faq}

\begin{itemize} 
\item[Q:]	Can we explain the preference for numbers of order one by arguing that such numbers more commonly appear in mathematics?
\item[A:]	No, in math you can find numbers of all sizes and shapes. It is true that the math-numbers we are exposed to in school -- $e$, $\gamma$, $\pi$, the Feigenbaum constants, and so on -- are of order 1. But if you dig around a little you find numbers both large and small in mathematics. A good example is the number of elements of the 16 sporadic groups which takes on values from $8\times10^3$ to about $10^{54}$.
	
\item[Q:]	Are you saying we should stop looking for explanations?
\item[A:]	Of course not. Any better explanation is a step forward. My point is that inventing a probability distribution to explain a parameter just adds unnecessary clutter. It doesn't explain anything and it's not good scientific practice.
	
\item[Q:]	Should finetuning arguments be discarded?
\item[A:]	Finetuning arguments work fine if one knowns the probability distribution. For example, we can make reasonable statements about how probable our galaxy or our solar system is because we have collected statistics from other galaxies and solar systems. 

The criticism of heliocentrism based on the argument that the absence of observable parallax implied the stars had to be ``unnaturally'' far away was wrong for exactly this reason: They had no probability distribution but erroneously postulated one by assuming that the stars should be likely to have similar distances to the planets as the planets have among each other. We now understand the distribution of stars and their typical distances comes about dynamically during structure formation and  that there is nothing ``unnatural'' about the distance of our Sun to the other suns. 

The relevance of the example from heliocentrism is that progress was not made by choosing the theory that ``naturally'' explained the absence of parallax by putting earth in the center of the universe. Instead, the correct explanation was that the small number was probable according to a suitable distribution. 

It would be possible to treat finetuning arguments as a hypotheses, but it doesn't presently look as if they're particularly well-working hypotheses. 
	
\item[Q:]	What should physicists do instead of obsessing about small numbers?
\item[A:]	I'd suggest they focus on well-defined problems, or at least make an effort to come up with well-defined problems. 

Given that the standard model is technically natural except for the Higgs-mass, it seems plausible that technical naturalness does have a rigorous mathematical basis under certain circumstances. The question is under which circumstances. 
	
\item[Q:]	Doesn't relying on Bayesian inference solve the problems with technical naturalness?
\item[A:] 	No, it doesn't. The Bayesian approach to technical naturalness is merely a different way to quantify the sensitivity of the low-energy parameters on the high-energy parameters. This is a good way to avoid having to pick one particular measure for naturalness. But I don't question the sensitivity itself; I question it is rational to believe a theory less sensitive to high energies is more likely to be correct. The Bayesian approach doesn't say anything about this. 

The reason that technically natural theories -- like eg supersymmetry -- come out ahead in Bayesian assessment is that these theories are more rigid in a well-defined sense: The additional symmetry (which is what makes the theory natural) favors more restrictive models because these models essentially have parameter-correlations built in by way of the symmetry requirement. 

These Bayesian assessments, however, do not quantify the presence of the additional axiom which is the symmetry itself (or whatever other assumption it is that makes a model natural). It is common practice in the literature of Bayesian assessments to compare models with different assumptions (rather than just the same model with different parameters), but that doesn't mean it's good practice. Of course if I add an assumption -- like supersymmetry -- which enforces a near-cancellation of parameters, and do not account for that assumption then this model will appear simpler and hence preferable. But this just moves the question from what was the probability distribution in theory-space to what were the priors of these different models to begin with. And there is no reason to assume that a theory is more likely to be a better description of nature just because it is more rigid. 
 \end{itemize}
 
\section{Why does it matter?}
\label{Why}

Physicists' belief that a correct theory should be natural was the reason many of them thought the LHC should see additional new particles besides the Higgs-boson. This has not happened. It is also the reason why dozens of experiments were commissioned to search for WIMPs, axions, and signals of a grand unification (like eg proton decay). We are here looking at billions of dollars of investment. 

While theoretical expectations are certainly not the only reason to commission experiments -- experiments are also driven by technological possibilities and experimentalists' interest -- it is without doubt one of the reasons. The focus on ill-motivated theories, therefore, creates a vicious cycle in which we attempt to find evidence for unpromising theories by experiments which deliver little guidance on the development of better theories, resulting in more fruitless theories and further experimental null-results. 

To showcase the concern, allow me to quote from a recent comment in {\sl Nature} by Ji Wu and Roger Bonnet \cite{17}.  In their comment, titled ``Maximize the impacts of space science,'' the two space scientists advocate that
\begin{quote}
``Agency managers should first assess options with the research community to reach a consensus on which scientific frontiers are most likely to yield major breakthroughs.''
\end{quote}

Naturalness arguments have been extremely important in the foundations of physics to quantify which ranges of parameter-space are promising to look for new phenomena. Their failure, therefore, requires attention and a revision of method.

Moreover, the continued failure of predictions in the fields of cosmology and particle physics erodes public trust in the foundations of physics. This is unfortunate because it is the area of science in which we are most likely to find entirely new laws of nature -- provided we look for evidence in the right places. 

\section{Conclusions}
\label{conc}

I have argued here that the popularity of arguments from naturalness and finetuning in the foundations of physics is problematic. These arguments are not mathematically well-defined because they refer to probabilities without a probability distribution. If one attempts to remove the problem by defining a probability distribution, this introduces an arbitrariness which conflicts with naturalness itself, because naturalness is a criterion invented to remove arbitrariness. If one does not specify the probability distribution (which is most often the case), naturalness remains ill-defined. Arguments from naturalness are then merely aesthetic criteria with little historical evidence of being useful. 
I conclude that attempts to solve ill-defined naturalness problems are a waste of time.

I will not deny that I too feel that finetuning is ugly and natural theories are more beautiful. I do not, however, see any reason for why this perception should be relevant for the discovery of more fundamental laws of nature. 

The title of this paper refers to a phrase I have heard unnaturally often, that certain coincidences ``scream for explanation.''

\section*{Acknowledgements}

I thank the Munich Center for Advanced Studies and the Munich Center for Mathematical Philosophy for hospitality and gratefully acknowledge support from the Foundational Questions Institute and the German Research Foundation.

\end{document}